\documentclass[11pt]{article}

\usepackage[final]{acl}

\usepackage{times}
\usepackage{latexsym}

\usepackage[T1]{fontenc}
\usepackage[utf8]{inputenc}

\usepackage{microtype}

\usepackage{inconsolata}

\usepackage{graphicx}
\usepackage{amsmath}
\usepackage{amssymb}
\usepackage{mathtools}
\usepackage{amsthm}
\usepackage{times}
\usepackage{latexsym}
\usepackage{booktabs}
\usepackage{tabularx}
\usepackage{multirow}
\usepackage{amsmath} 
\usepackage{graphicx}
\usepackage{adjustbox}
\usepackage{pdflscape}

\title{ The Limits of LLM Forecasting: \\Parametric Knowledge Gaps Across Conflict Zones}

\author{Poli Nemkova \\
  University of North Texas, 
  Department of Computer Science and Engineering\\
  \texttt{poli.nemkova@unt.edu} 
}
\begin{document}
\maketitle
\begin{abstract}
Media coverage of armed conflict is deeply asymmetric: we document a 
224$\times$ gap between the most and least covered conflict zones in 
English-language media across 22 countries (2020--2026). We evaluate 
zero-shot conflict escalation forecasting across all 22 countries on 
a 660-case held-out test set, comparing Llama-3.3-70B and GPT-4o 
against three structured baselines.

The central finding is not a performance gradient but a qualitative 
failure: LLMs do not forecast conflict---they categorize it. Llama 
predicts escalation on every under-covered case, matching the trivial 
Always-YES baseline to three decimals; GPT-4o predicts NO on every 
over-covered case, missing all five actual escalation events. A 
logistic regression using only eleven observation-window features with 
\emph{no country information} achieves F1~=~0.402, outperforming both 
LLMs in every measurable tier. This failure cannot be resolved at 
inference time: adding structured ACLED evidence degrades performance 
on under-covered zones (GPT-4o F1: 0.323~$\to$~0.168) and falls 
below LR by a factor of 2.4. The bottleneck is not data availability 
but the LLM's interpretation of temporal signal under a 
country-categorical prior.

Under-covered populations receive not just less accurate AI, but 
qualitatively different AI that cannot distinguish stable from 
escalating periods. We call for coverage-stratified benchmarking, 
conflict NLP datasets for under-covered zones, and training data 
documentation standards for geographic conflict representation.

\end{abstract}

\section{Introduction}
\begin{figure*}[t]
  \centering
  \includegraphics[width=2.1\columnwidth]{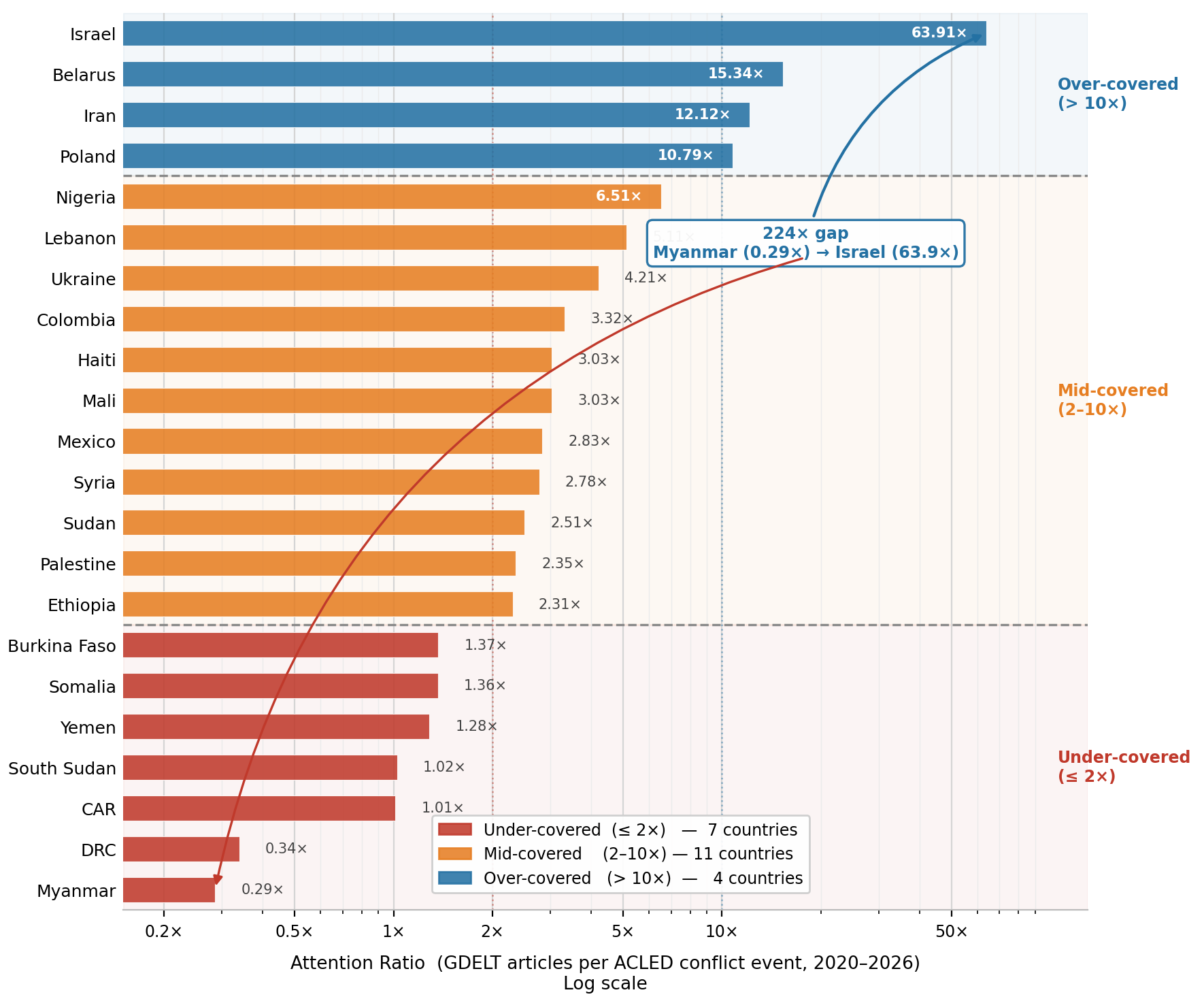}
  \caption{Media attention ratio across 22 conflict zones, measured as
  GDELT English-language news articles per ACLED conflict event,
  2020--2026 (log scale). A 224$\times$ gap separates the most and least
  covered zones.}
  \label{fig:attention_gap}
\end{figure*}

AI systems for humanitarian decision support increasingly rely on LLMs
for conflict early warning, displacement forecasting, and resource
allocation \citep{karamolegkou-etal-2026-nlp, rost2026anticipating, unhcr2022jetson}.
Yet these systems inherit world knowledge from training corpora shaped
by highly uneven English-language media coverage. This paper asks
whether media attention asymmetries translate into asymmetries in how
LLMs reason about conflict, and what this means for humanitarian AI.

Western media bias in conflict and disaster coverage is well documented.
Prior work shows that proximity, cultural consonance, elite-nation
involvement, and framing shape which crises become salient
\citep{galtung1965structure, entman1993framing}. Quantitative studies
further show that news attention affects real-world aid allocation:
\citet{eisensee2007news} found that African disasters require far more
casualties than comparable Eastern European disasters to receive similar
US television coverage. Recent computational work confirms persistent
under-coverage of conflicts in Sub-Saharan Africa and Southeast Asia
relative to Europe and the Middle East, even after accounting for
conflict intensity \citep{croicu2024combining}. What remains unknown is
whether such asymmetries propagate into LLM-based conflict reasoning.

This question matters because LLMs are increasingly considered as
reasoning engines for forecasting, while prior conflict prediction work
has mostly relied on structured datasets and hand-engineered features
\citep{ward2013learning, blair2020forecasting}. At the same time, NLP
research shows that models inherit biases from web-scale corpora
\citep{bender2021dangers, dodge2021documenting, gebru2021datasheets},
including geographic bias in prediction tasks
\citep{manvi2024geographically}, multilingual instability
\citep{wu2020languages, nemkova2025cross}, and downstream failures in
humanitarian contexts \citep{nemkova2025comparing}. However, geographic
conflict coverage asymmetry has not been studied as a source of
qualitative reasoning failure.

We make five contributions. First, using ACLED event data
\citep{raleigh2010acled} and GDELT media coverage
\citep{leetaru2013gdelt} from 2020--2026, we document a
\textbf{224$\times$ English-language media attention gap} across 22 active
conflict zones, from Myanmar at 0.29 articles per event to Israel at
63.9.

Second, we show that this asymmetry produces not merely unequal accuracy,
but different reasoning modes. On a 660-case held-out test set, Llama
predicts escalation for all 210 under-covered cases, matching an
Always-YES baseline, while GPT-4o predicts no escalation for all 120
over-covered cases, missing every true escalation event. The models do
not forecast conflict dynamics; they categorize conflict contexts.

Third, we show that the task is forecastable from temporal signal alone:
a logistic regression using only eleven observation-window features and
no country information achieves F1 = 0.402, outperforming both LLMs in
their parametric condition. This indicates that LLM failure stems from
country-level categorical priors overriding available temporal evidence.

Fourth, structured evidence at inference time does not fix the problem.
Adding ACLED event counts degrades LLM performance on under-covered
zones, with GPT-4o-evidence falling to F1 = 0.168, far below logistic
regression. The bottleneck is therefore not data availability, but how
LLMs interpret temporal evidence under geographic priors.

Fifth, we argue that this is a first-class fairness problem: under-covered
populations receive not only less accurate AI, but qualitatively different
AI that cannot distinguish stable from escalating periods within their
conflicts. We call for coverage-stratified benchmarking, stronger conflict
NLP datasets for under-covered regions, and training-data documentation
standards for geographic conflict representation.

\section{Measuring the Gap}
\label{gap}
\subsection{Data Sources}

We operationalize media attention using two independent data sources that together
provide ground-truth conflict intensity and media coverage volume across 22 active
conflict zones from January 2020 through February 2026.

\textbf{Conflict ground truth: ACLED.} The Armed Conflict Location and Event Dataset
\citep{raleigh2010acled} provides event-level records of political violence and
protest events, coded from news reports, NGO communications, and government sources.
We use ACLED as our measure of actual conflict intensity, aggregating monthly event
counts per country. ACLED's systematic global coverage and real-time coding make it
the standard ground truth for conflict research \citep{muchlinski2016comparing}.
Crucially, ACLED itself is subject to reporting bias — events in countries with less
media access are systematically under-coded \citep{raleigh2010acled} — meaning our
attention gap estimates are \textit{conservative}: the true asymmetry between conflict
intensity and media coverage is likely larger than we measure.

\textbf{Media coverage: GDELT.} The Global Database of Events, Language and Tone
\citep{leetaru2013gdelt} indexes English-language news articles from thousands of
outlets worldwide, coding events using the CAMEO conflict taxonomy. We query the
GDELT BigQuery dataset for conflict-relevant CAMEO event codes (codes 14--20, 173,
180--196, 200--203) and aggregate monthly counts of \textit{distinct source URLs}
per country. URL-level deduplication is essential: each news event in GDELT generates
multiple rows (one per actor pair, one per event code), so raw row counts massively
overcount article volume. We use distinct URLs as the closest available proxy for
distinct articles. We note that GDELT's coverage is substantially English-language
biased, systematically under-indexing local-language reporting on conflicts in
Sub-Saharan Africa and Southeast Asia. This is not a limitation we correct for —
it is precisely the bias we seek to measure, since English-language web text
constitutes a substantial portion of LLM pretraining corpora
\citep{dodge2021documenting}.

\textbf{Country selection.} We select 22 countries spanning five regions: Sub-Saharan
Africa (Sudan, Ethiopia, Somalia, DRC, South Sudan, Nigeria, Mali, Burkina Faso,
Central African Republic), Southeast Asia (Myanmar), Eastern Europe (Ukraine, Belarus,
Poland), Middle East and North Africa (Israel, Palestine, Lebanon, Iran, Yemen, Syria),
and Latin America (Colombia, Haiti, Mexico). Countries were selected to provide
structural diversity across conflict types — interstate war, post-coup civil conflict,
protracted insurgency, multi-actor fragmented conflict, and political crisis — while
ensuring sufficient ACLED event density for meaningful monthly statistics. Countries
with near-zero ACLED event counts (Belarus, Poland) are retained as implicit controls
for the model behavior analysis in Section~\ref{sec:llm}.

\subsection{The Attention Ratio}

We define the \textit{media attention ratio} for country $c$ as:

\begin{equation}
\text{AttentionRatio}(c) = 
\frac{\overline{A}_{\text{GDELT}}(c)}
     {\overline{E}_{\text{ACLED}}(c)}
\end{equation}

where both quantities are averaged over the 74-month study period. This ratio
captures media attention \textit{per unit of conflict activity}: a high ratio
indicates that each conflict event generates many articles (over-covered); a low
ratio indicates that conflict events generate few articles (under-covered).

\subsection{Results: A 224$\times$ Attention Gap}

Figure~\ref{fig:attention_gap} reports the attention ratio for all 22 countries, sorted
from least to most covered. The gap between the most and least covered active conflict
zones spans \textbf{224$\times$}: Myanmar receives 0.29 articles per conflict event
while Belarus receives 15.34$\times$ and Israel 63.9$\times$. Restricting to countries
with substantial active conflict (ACLED events $>$ 1,000 over the study period)
narrows the comparison but preserves the essential finding: Myanmar (0.29$\times$)
and South Sudan (1.02$\times$) receive approximately \textbf{15--220$\times$ less
media attention per conflict event} than Ukraine (4.21$\times$), Lebanon
(6.51$\times$), or Iran (12.12$\times$).

Three structural patterns are visible. First, the seven most under-covered active
conflict zones are without exception located in Sub-Saharan Africa or Southeast Asia:
Myanmar, DRC, South Sudan, CAR, Yemen, Somalia, and Burkina Faso. Second, the most
over-covered zones — Israel, Belarus, Iran, Poland — share a common feature: their
media coverage reflects geopolitical salience rather than conflict intensity.  Poland's coverage reflects spillover from the Ukraine
refugee crisis rather than domestic armed conflict. Third, Ukraine's ratio (4.21$\times$)
falls in the mid-covered tier despite being the largest absolute recipient of media
attention in our dataset (1.1 million articles), because its ACLED event count is
also the largest (266,664 events). High absolute coverage does not necessarily imply
high relative coverage.

\subsection{Methodological Notes}

Two limitations deserve explicit acknowledgment. First, GDELT indexes primarily
English-language sources. Local-language reporting on conflicts in Bambara, Hausa,
Tigrinya, or Burmese is substantially absent, meaning our attention ratios understate
the true coverage available to locally-informed observers. Since LLM pretraining
corpora also predominantly reflect English-language sources \citep{dodge2021documenting},
this shared bias is precisely what we seek to characterize rather than correct.
Second, the attention ratio captures volume but not quality of coverage: a country
may receive many articles but superficial treatment, or few articles but detailed
analysis. We treat volume as a proxy for training signal density, acknowledging that
this is an approximation. Future work employing document-level analysis of coverage
depth could refine this measure.

\section{Experimental Setup}

\label{sec:setup}

We evaluate zero-shot conflict escalation forecasting across all 22
countries using two LLM backbones: Llama-3.3-70B \citep{grattafiori2024llama}
served via the Groq inference API, and GPT-4o \citep{openai2023gpt4}.
The zero-shot parametric design is deliberate: we provide only the country
name and prediction period, with no evidence bundle, no few-shot examples,
and no contextual data. Any performance difference across coverage tiers
must therefore reflect parametric knowledge — what each model learned
during pretraining — rather than evidence quality or prompting strategy.

The zero-shot design follows the evaluation paradigm established
in \citet{brown2020language}: by withholding in-context examples,
we isolate what the model has internalized during pretraining
from what it can infer from prompt context. This is the
appropriate design for our research question, which concerns
parametric knowledge rather than in-context reasoning.

The task is binary escalation forecasting: given a country and a 30-day
prediction window, predict whether armed conflict fatalities will exceed
130\% of the recent baseline. Ground truth is derived from ACLED as
described in Section~\ref{gap}. We evaluate across the full
1,628-case dataset (74 cases per country, 2020--2026), using temperature
$T=0$ for deterministic output and a single run per model.

We compare against two baselines. The \textbf{majority-class baseline}
predicts the most frequent label per country; because all 22 countries
have escalation rates below 0.5, this always predicts NO, yielding
F1 = 0.000 everywhere and providing no discriminative signal. The
\textbf{Always-YES baseline} predicts escalation on every case,
maximizing recall at the cost of precision. Because the majority-class
baseline is trivially uninformative, Always-YES serves as the operative
comparison: it represents the simplest possible strategy for catching
escalation events in chronically conflicted zones, and any claimed
improvement should be measured against it.

To test whether structured evidence can close the performance gap on
under-covered zones, we run a second condition (\textbf{evidence-augmented})
in which each prompt includes the three-month ACLED event count trend
preceding the prediction window. This ablation directly addresses the
hypothesis that the parametric condition's failures stem from insufficient
information rather than miscalibrated priors. We report evidence-augmented
results for both models in Section~\ref{sec:ablation}.

\section{LLM Behavior Under Attention Asymmetry}
\label{sec:llm}

\subsection{Establishing the Signal: Structured Baselines}
\label{subsec:structured_baselines}

Before evaluating LLM behavior, we establish that the task contains
exploitable temporal signal. If simple structured models using only
observation-window features can outperform trivial baselines, then any
LLM failure on the same task cannot be attributed to task difficulty —
it must reflect a failure to use information that is empirically
present in the data.

We evaluate three structured baselines on a per-country chronological
train/test split of the 1{,}628-case dataset: the earliest 60\% of
months per country are used for training (968 cases, 178 positives) and
the most recent 40\% are held out for evaluation (660 cases, 132
positives; per-tier positives are 53 under-covered, 74 mid-covered, and
5 over-covered). All conditions in this paper — structured and LLM
alike — are evaluated on the same 660-case held-out test set, so all
F1 comparisons are case-aligned.

\textbf{Always-YES} predicts escalation on every case.
\textbf{Persistence} predicts escalation if the observation window's
recency ratio exceeds 1.3 — that is, if fatalities in the last 14 days
of the observation window were more than 1.3$\times$ the fatalities of
the prior 14 days. This is the same 1.3$\times$ threshold used to define
the label, applied as an "is violence already accelerating" rule using
only past data.
\textbf{Logistic regression (LR)} uses eleven pre-computed
observation-window features: eight numeric aggregates (total fatalities,
event count, maximum event fatalities, average daily fatalities,
fatalities per event, volatility, 14-day recency ratio, and 2-week
baseline) plus a one-hot encoding of the trend category
(\textit{stable}, \textit{increasing}, \textit{decreasing}). The LR
receives no country identity information of any kind: no country name,
no country fixed effects, no coverage tier feature. Its predictions are
functions only of the observation window's internal temporal structure.
Hyperparameters are class-balanced L2-regularized logistic regression
with $C=1.0$, standard-scaled features.

Table~\ref{tab:structured_baselines} reports tier-stratified F1.
\textbf{Logistic regression using only observation-window features,
with no country information, outperforms Always-YES overall
(F1 = 0.402 vs.\ 0.333) and in two of three tiers}, with the only
exception being a 0.003-point gap on under-covered (LR = 0.400 vs.\
Always-YES = 0.403) where both achieve similar F1 through very different
precision-recall profiles. The LR's Spearman correlation between
attention ratio and F1 is $\rho = -0.019$ ($p = 0.94$), essentially
zero — confirming that country-blind temporal features yield
country-invariant performance.

\begin{table}[t]
\centering
\small
\begin{tabular}{@{}lrrr@{}}
\toprule
\textbf{Condition} & \textbf{Under} & \textbf{Mid} & \textbf{Over} \\
\midrule
Always-YES        & \textbf{0.403} & 0.366 & 0.080 \\
Persistence       & 0.362 & 0.273 & 0.231 \\
LR (no country)   & 0.400 & \textbf{0.408} & \textbf{0.370} \\
\bottomrule
\end{tabular}
\caption{Structured-baseline F1 by coverage tier. Held-out test set:
660 cases, 132 positives (53 under, 74 mid, 5 over). Per-country
chronological 60/40 split of the 1{,}628-case dataset; LR is trained
on the earliest 60\% of each country's history. LR uses eleven
observation-window features and \emph{no country information}.}
\label{tab:structured_baselines}
\end{table}

These results reframe the evaluation question for the LLM conditions.
The signal is exploitable. The signal does not require country identity.
The question becomes not whether LLMs can forecast under uncertainty,
but whether they can extract from country names and parametric knowledge
the same information that eleven feature values encode directly.

\subsection{Results: Two Models, Two Failure Modes}
\label{subsec:two_failures}

Table~\ref{tab:parametric_results} reports per-tier performance for
both LLMs alongside the three structured baselines, all evaluated on
the same 660-case held-out test set. The headline finding is not a
gradient of forecasting skill correlated with attention ratio — it is
a qualitative divergence in how each model handles uncertainty about
unfamiliar conflict zones, relative to baselines that exploit temporal
signal without country identity.

\begin{table*}[t]
\centering
\small
\begin{tabular*}{\textwidth}{@{\extracolsep{\fill}} l l r r r r @{}}
\toprule
\textbf{Condition} & \textbf{Tier} & \textbf{F1} & \textbf{P} & \textbf{R} & \textbf{N+} \\
\midrule
\multirow{3}{*}{\normalsize Always-YES}
  & Under  & 0.403 & 0.252 & 1.000 & 53 \\
  & Mid    & 0.366 & 0.224 & 1.000 & 74 \\
  & Over   & 0.080 & 0.042 & 1.000 &  5 \\
\midrule
\multirow{3}{*}{\normalsize Persistence}
  & Under  & 0.362 & 0.333 & 0.396 & 53 \\
  & Mid    & 0.273 & 0.292 & 0.257 & 74 \\
  & Over   & 0.231 & 0.143 & 0.600 &  5 \\
\midrule
\multirow{3}{*}{\normalsize LR (no country info)}
  & Under  & 0.400 & 0.291 & 0.641 & 53 \\
  & Mid    & \textbf{0.408} & 0.333 & 0.527 & 74 \\
  & Over   & \textbf{0.370} & 0.227 & 1.000 &  5 \\
\midrule
\multirow{3}{*}{\normalsize Llama-3.3-70B (parametric)}
  & Under  & 0.403 & 0.252 & 1.000 & 53 \\
  & Mid    & 0.386 & 0.261 & 0.743 & 74 \\
  & Over   & 0.100 & 0.057 & 0.400 &  5 \\
\midrule
\multirow{3}{*}{\normalsize GPT-4o (parametric)}
  & Under  & 0.323 & 0.221 & 0.604 & 53 \\
  & Mid    & 0.356 & 0.281 & 0.486 & 74 \\
  & Over   & 0.000 & 0.000 & 0.000 &  5 \\
\bottomrule
\end{tabular*}
\caption{Zero-shot escalation forecasting by coverage tier on the
660-case held-out test set (per-country chronological 60/40 split).
Coverage tiers are defined by attention ratio: Under
($\leq$2$\times$), Mid (2--10$\times$), and Over ($>$10$\times$). LR
uses no country information; both LLMs receive country name and
prediction window. N+ denotes positive cases per tier (53 under, 74 mid,
5 over). LLM results are filtered from the original 1{,}628-case
evaluation to the same 660 case IDs as the structured baselines, so all
F1 values are directly comparable.}
\label{tab:parametric_results}
\end{table*}

\textbf{Finding 1: Llama under-covered behavior is exactly Always-YES.}
Llama achieves under-covered F1 = 0.403, precision = 0.252,
recall = 1.000 — values that are identical to the Always-YES baseline
to three decimal places. On 210 under-covered cases spanning seven
countries (Burkina Faso, Central African Republic, DRC, Myanmar,
Somalia, South Sudan, Yemen) and six years of monthly observations,
Llama predicts escalation on every single case. A frontier 70-billion
parameter model trained on a substantial portion of the public web
reduces, on the world's most neglected conflict zones, to the trivial
predictor that says "yes" to everything. This is the cleanest possible
empirical instantiation of categorization-rather-than-forecasting.

\textbf{Finding 2: GPT-4o predicts NO on every over-covered case.}
GPT-4o achieves over-covered F1 = 0.000, recall = 0.000. Across 120
cases in Belarus, Iran, Israel, and Poland, GPT-4o does not predict
escalation once — missing all five actual escalation events, including
the October 2023 escalation in Israel. This is the mirror image of
Llama's failure on under-covered zones: where Llama applies a universal
YES prior, GPT-4o applies a universal NO prior. More capable parametric
knowledge produces more confident miscategorization, not better
discrimination.

\textbf{Finding 3: Structured models outperform LLMs on every tier
where positives exist.}
LR with eleven observation-window features and no country information
achieves F1 = 0.402 overall — exceeding both Llama (0.374) and GPT-4o
(0.333). The contrast is sharpest on over-covered zones, where LR
achieves F1 = 0.370 with recall = 1.000 — catching all 5 escalation
events that GPT-4o misses entirely. On mid-covered zones LR also
exceeds both LLMs (0.408 vs.\ Llama 0.386 vs.\ GPT-4o 0.356). On
under-covered zones, LR (0.400), Llama (0.403), and Always-YES (0.403)
are essentially tied in F1 — but the precision-recall profiles differ
sharply: LR achieves P = 0.291, R = 0.641, the only condition with
balanced discrimination, while Llama and Always-YES both achieve their
F1 through degenerate recall = 1.000 predictions.

\textbf{Finding 4: The task is not intractable; LLMs fail to use signal
that eleven features encode.}
LR has \emph{less} information than either LLM: no country name, no
world knowledge of conflict actors, no political context. It receives
only eleven numeric and categorical aggregates over a 30-day
observation window. Yet it produces a deployable forecast: balanced
precision and recall, country-invariant performance (Spearman
$\rho = -0.019$), and the only F1 above 0.10 on over-covered zones.
The LLMs, despite having every piece of information LR has plus full
parametric knowledge of conflict history, fail to extract from country
names what eleven observation-window numbers encode directly.

\section{Ablation: Does Evidence Help?}
\label{sec:ablation}

The parametric results in Section~\ref{sec:llm} leave open a natural
objection: perhaps LLM failure on under-covered zones reflects not a
fundamental knowledge deficit but simply an absence of real-time
signal. If so, providing structured conflict evidence at inference
time should close the gap. We test this directly by augmenting each
prompt with the three-month ACLED event count trend preceding the
observation window, and re-evaluating both LLMs on the same 660-case
held-out test set.

\subsection{Results}
\label{subsec:evidence_results}

Table~\ref{tab:five_condition} reports the full seven-condition
comparison: three structured baselines (Always-YES, Persistence, LR),
both LLMs in the parametric condition, and both LLMs in the
evidence-augmented condition. Three findings stand out.

\begin{table*}[t]
\centering
\small
\begin{tabular*}{\textwidth}{@{\extracolsep{\fill}} l l r r r r @{}}
\toprule
\textbf{Condition} & \textbf{Tier} & \textbf{F1} & \textbf{P} & \textbf{R} & \textbf{N+} \\
\midrule
\multirow{3}{*}{\normalsize Always-YES}
  & Under & 0.403 & 0.252 & 1.000 & 53 \\
  & Mid   & 0.366 & 0.224 & 1.000 & 74 \\
  & Over  & 0.080 & 0.042 & 1.000 &  5 \\
\midrule
\multirow{3}{*}{\normalsize Persistence}
  & Under & 0.362 & 0.333 & 0.396 & 53 \\
  & Mid   & 0.273 & 0.292 & 0.257 & 74 \\
  & Over  & 0.231 & 0.143 & 0.600 &  5 \\
\midrule
\multirow{3}{*}{\normalsize LR (no country info)}
  & Under & 0.400 & 0.291 & 0.641 & 53 \\
  & Mid   & \textbf{0.408} & 0.333 & 0.527 & 74 \\
  & Over  & \textbf{0.370} & 0.227 & 1.000 &  5 \\
\midrule
\multirow{3}{*}{\normalsize Llama-3.3-70B (parametric)}
  & Under & \textbf{0.403} & 0.252 & 1.000 & 53 \\
  & Mid   & 0.386 & 0.261 & 0.743 & 74 \\
  & Over  & 0.100 & 0.057 & 0.400 &  5 \\
\midrule
\multirow{3}{*}{\normalsize GPT-4o (parametric)}
  & Under & 0.323 & 0.221 & 0.604 & 53 \\
  & Mid   & 0.356 & 0.281 & 0.486 & 74 \\
  & Over  & 0.000 & 0.000 & 0.000 &  5 \\
\midrule
\multirow{3}{*}{\normalsize Llama-3.3-70B (evidence)}
  & Under & 0.342 & 0.300 & 0.396 & 53 \\
  & Mid   & 0.212 & 0.188 & 0.243 & 74 \\
  & Over  & 0.150 & 0.086 & 0.600 &  5 \\
\midrule
\multirow{3}{*}{\normalsize GPT-4o (evidence)}
  & Under & 0.167 & 0.186 & 0.151 & 53 \\
  & Mid   & 0.165 & 0.155 & 0.176 & 74 \\
  & Over  & 0.188 & 0.111 & 0.600 &  5 \\
\bottomrule
\end{tabular*}
\caption{Seven-condition comparison on the 660-case held-out test set.
Evidence condition augments each prompt with the three-month ACLED
event count trend preceding the observation window. All conditions
evaluated on identical cases. N+ = positive cases per tier.}
\label{tab:five_condition}
\end{table*}

\textbf{Finding 1: Adding evidence degrades LLM performance everywhere
except on the over-covered tier.} Reading down each LLM's column from
parametric to evidence: Llama under-covered F1 falls from 0.403 to
0.342; Llama mid-covered F1 falls from 0.386 to 0.212; GPT-4o
under-covered F1 falls from 0.323 to 0.167; GPT-4o mid-covered F1
falls from 0.356 to 0.165. Overall, Llama drops from F1 = 0.374 to
0.252 and GPT-4o drops from 0.333 to 0.168. The only places evidence
helps are tiers where the parametric prior had collapsed entirely:
GPT-4o over-covered moves from F1 = 0.000 to 0.188 (recall from 0.000
to 0.600), and Llama over-covered moves from 0.100 to 0.150. Evidence
helps precisely where the model had nothing to begin with, and hurts
everywhere else.

\textbf{Finding 2: Evidence augmentation falls below both Always-YES
and structured baselines on under-covered zones.}
On the populations the paper is centrally about, both evidence
conditions are dominated by the trivial Always-YES predictor
(F1 = 0.403) and by LR (F1 = 0.400): Llama-evidence achieves F1 = 0.342,
GPT-4o-evidence F1 = 0.167. A humanitarian practitioner who replaced
the trivial baseline with either evidence-augmented model would catch
fewer escalation events in the world's most neglected conflict zones.
The mechanism is interpretable. When presented with event counts, both
LLMs attempt to reason about whether the observed trend constitutes
meaningful escalation --- but they lack country-specific calibration
to do so. A three-month count of 200 ACLED events means something very
different in Myanmar (chronically high-intensity) than in South Sudan
(episodic). The parametric prior --- blanket YES for conflict zones,
in Llama's case --- is disrupted by the new input, but the LLM does
not develop a calibrated trend interpretation to replace it.

\textbf{Finding 3: The same data, processed by logistic regression,
beats every LLM evidence condition by a factor of two.} The clearest
diagnostic in the table is the direct comparison between LR
(F1 = 0.402 overall) and GPT-4o-evidence (F1 = 0.168 overall). Both
have access to ACLED-derived numerical features over a recent window
preceding the prediction period. LR receives eleven features as a
scaled vector and produces a balanced precision-recall forecast across
all three tiers, including F1 = 0.370 with recall = 1.000 on
over-covered zones. GPT-4o-evidence receives three event counts as
text, plus full parametric world knowledge of every country in the
dataset, and produces F1 below 0.20 in every tier. The information
bottleneck is not the data; it is the LLM's interpretation of it.

This asymmetry confirms that the failure on under-covered zones is a
parametric-knowledge problem rather than a prompting problem. Evidence
augmentation is not a viable fix at inference time: it improves
performance only where parametric knowledge had completely failed (the
over-covered tier, where GPT-4o predicts NO on every case in the
parametric condition), and worsens it across the populations where the
humanitarian stakes are highest.

\subsection{Implications for System Design}
\label{subsec:system_design}

These results carry a direct implication for practitioners building
LLM-assisted humanitarian early warning systems. The standard design
assumption --- that richer prompts produce better forecasts --- does
not hold across coverage tiers, and inverts on the populations that
matter most. On under-covered zones, the parametric Llama model with
no evidence (F1 = 0.403) outperforms both evidence-augmented LLMs and
matches the trivial Always-YES baseline; on mid-covered zones, the
parametric models outperform their own evidence-augmented variants;
only on over-covered zones --- where the parametric prior had already
collapsed to recall = 0.000 for GPT-4o --- does adding evidence help.

The appropriate benchmark for any proposed LLM-based early warning
system is not Always-YES on well-covered zones; it is a logistic
regression on observation-window features in the geographies where the
system is most likely to be deployed. None of the four LLM conditions
tested here clear that bar on under-covered or mid-covered tiers, and
the evidence-augmented conditions do not clear it on any tier. The
implication is not that LLMs cannot be useful for humanitarian
forecasting --- it is that current LLM deployment patterns assume an
ability to extract structured temporal signal from natural-language
prompts that frontier models do not yet exhibit on the geographic
contexts where humanitarian deployment is most consequential.

\section{Recommendations}
\label{sec:recommendations}

Our findings motivate three concrete recommendations, framed as
actionable research directions with clear precedent.

\subsection{Coverage-Stratified Benchmarking}

Conflict NLP evaluations should report performance stratified by
media attention ratio alongside aggregate metrics. A system that
performs well on Ukraine and Palestine while failing on DRC and
Myanmar is not a general-purpose conflict AI system --- it is
calibrated to well-covered conflicts. We propose a simple standard:
report F1 separately for under-covered ($\leq$2$\times$),
mid-covered (2--10$\times$), and over-covered ($>$10$\times$) tiers
using the attention ratios we provide. This requires no additional
data collection --- only a change in reporting practice, analogous
to the per-language evaluation norms that exposed systematic
disparities in multilingual NLP \citep{wu2020languages}.

\subsection{Dataset Creation for Under-Covered Conflict Zones}

The training signal deficit for DRC, Myanmar, South Sudan, and
similar contexts cannot be addressed by better architectures or
prompting alone --- our evidence condition results confirm this
directly. We call on the NLP community to treat annotated conflict
datasets for under-covered zones as priority infrastructure:
creating NLP-ready resources from ACLED and UCDP \citep{sundberg2013introducing} for under-covered
regions, supporting annotation projects involving local researchers,
and funding shared tasks targeting these contexts --- the same
deliberate investment that drove progress in low-resource MT
\citep{wu2020languages}. The data infrastructure exists. What is
missing is the community's attention to it.

\subsection{Training Data Documentation}

LLM developers should document geographic conflict coverage
statistics --- specifically, what fraction of training data
discusses each conflict zone and how this correlates with
task performance. This instantiates the agenda of
\citet{gebru2021datasheets} and \citet{dodge2021documenting}
in a domain where underdocumented bias has measurable humanitarian
consequences. GPT-4o's recall of 0.143 on Somalia is not
discoverable from any model card; it requires empirical evaluation
most humanitarian organizations cannot conduct. We do not claim
developers must equalize coverage --- only that asymmetry, where
it exists, should be documented so deployers can make informed
decisions.

\section{Conclusion}
\label{sec:conclusion}
LLMs do not forecast conflict --- they categorize it. A logistic
regression with eleven observation-window features and no country
information achieves F1~=~0.402, outperforming both LLMs in every
condition. On 210 under-covered cases, Llama matches the trivial
Always-YES baseline to three decimals; on 120 over-covered cases,
GPT-4o misses every escalation event. Adding structured evidence
worsens this: the same ACLED data processed by logistic regression
beats GPT-4o-evidence by a factor of 2.4. The bottleneck is not
data availability but the LLM's interpretation of temporal signal
under a country-categorical prior.

The communities most affected by conflict and least covered by media
receive AI that cannot distinguish a stable month from an escalating
one --- not because the signal is absent, but because LLMs do not
use it. Closing this gap requires coverage-stratified benchmarks,
training datasets for under-covered zones, and documentation standards
that tell deployers where a model can and cannot be trusted.

\section*{Limitations}
We identify eight limitations of the present study.

\textbf{GDELT as an attention proxy.} GDELT indexes primarily
English-language outlets, underrepresenting local-language reporting.
A conflict zone with rich Arabic, French, or Swahili coverage will
appear under-covered in our measure even if it is not under-covered
globally. Our attention ratios are therefore conservative estimates:
the true asymmetry in LLM training data is likely larger than we
report. Additionally, GDELT aggregates across all news types, not
conflict-specific reporting, meaning politically salient countries
receive inflated ratios relative to their actual conflict coverage.

\textbf{ACLED coverage and conflict definition.} ACLED's coverage is
not uniform --- under-covered countries may have thinner event records
because ACLED's own documentation infrastructure is less developed
there, not because conflict is less frequent. Our escalation threshold
(130\% of a 2-week rolling fatality baseline) is operationally
motivated but arbitrary; absolute F1 values are threshold-dependent
even if the qualitative tier-level pattern is not.

\textbf{Two LLMs may not generalize.} We evaluate two frontier model
families. The categorical reasoning pattern we document may not hold
for smaller models, models with more recent training cutoffs, models
trained on conflict-specific corpora, or reasoning-augmented variants
(o1/o3, Claude with extended thinking). Our claim is that the pattern
exists in two widely-deployed systems that are plausible candidates
for humanitarian AI deployment; extending the evaluation across model
families and reasoning architectures is important future work.

\textbf{Evidence representation.} Our evidence-augmented condition
provides three monthly ACLED event counts as text, reflecting
realistic real-time data availability (fatality counts often lag by
days or weeks while event counts are produced first). The logistic
regression baseline uses a richer feature set including both event
counts and fatality aggregates, and beats every LLM condition by a
factor of two. We cannot rule out that LLMs given fatality-augmented
or differently-structured evidence would perform better; we leave
richer evidence representations as future work. We note, however, that
the structured-baseline contrast already establishes that the failure
is not about the kind of evidence available --- it is about the
language model's capacity to extract a forecasting signal from a
numerical text input that a non-language model extracts from features.

\textbf{Evidence timing.} Our evidence covers the three months
preceding the observation window, matching the temporal stance of a
real-time forecaster who works with finalized monthly aggregates. A
reviewer might ask whether LLMs would do better with observation-window
event counts directly; we leave this comparison to future work, while
noting that the logistic regression baseline has access to
observation-window features and still beats LLM-evidence by a wide
margin.

\textbf{Zero-shot evaluation only.} Few-shot prompting with
conflict-specific examples or retrieval-augmented generation may
improve performance on under-covered zones. We regard this as an open
question: the evidence-condition results suggest that adding structured
data does not fix the calibration problem, but more sophisticated
inference-time strategies remain untested.

\textbf{Training cutoff confounds.} Our dataset spans 2020--2026 and
both models have training cutoffs within this window. We do not
stratify by pre- and post-cutoff periods. However, the stable YES or
NO priors we observe across the full 74-month window --- and the
empirical identity between Llama's under-covered behavior and the
trivial Always-YES baseline (F1, precision, and recall all matching to
three decimals) --- are inconsistent with a model making genuine use
of temporal information; this pattern is most parsimoniously explained
by parametric-knowledge structure rather than by event memorization
for specific time periods.

\textbf{Causal interpretation.} We document a correlation between
media attention ratio and LLM behavior and interpret it as a training
data exposure effect. We cannot establish this causally --- we have no
access to either model's training data distribution. The alternative
explanation (that under-covered countries are harder to forecast due
to ACLED sparsity rather than LLM pretraining gaps) is argued against
by the structured-baseline results: a logistic regression using the
same ACLED data --- with no country information --- achieves
F1 = 0.402 overall, including F1 = 0.400 on under-covered zones. If
ACLED sparsity were the cause of LLM failure, LR would fail at the
same rate. It does not.

\textbf{Potential Risks. }This work could be misread as discouraging LLM use in humanitarian contexts entirely. Our findings are specific to zero-shot parametric forecasting; we do not evaluate fine-tuned, few-shot, or retrieval-augmented systems. Practitioners should not generalize our negative results beyond the conditions tested.

\bibliography{custom}

\appendix
\clearpage
\onecolumn

\section{Per-Country Results}
\label{sec:appendix}

\begin{center}
\footnotesize
\setlength{\tabcolsep}{3.5pt}
\renewcommand{\arraystretch}{1.02}

\captionof{table}{Per-country escalation forecasting results on the 660-case held-out test set: parametric condition. N\textsuperscript{+} = positive cases. Italicized rows show tier-level aggregates matching Tables~2 and~3. In the parametric condition, Llama achieves R = 1.000 on all seven under-covered countries, confirming the Always-YES pattern; GPT-4o achieves R = 0.000 on all four over-covered countries and four mid-covered countries.}
\label{tab:per_country_parametric}

\vspace{4pt}

\begin{adjustbox}{max width=\textwidth,center}
\begin{tabular}{@{}llrrrrrrrr@{}}
\toprule
& & & & \multicolumn{3}{c}{Llama-3.3-70B} & \multicolumn{3}{c}{GPT-4o} \\
\cmidrule(lr){5-7} \cmidrule(lr){8-10}
Tier & Country & N & N\textsuperscript{+} & F1 & P & R & F1 & P & R \\
\midrule
Under & Burkina Faso  & 30 & 10 & .500 & .333 & 1.000 & .500 & .333 & 1.000 \\
Under & CAR           & 30 &  9 & .462 & .300 & 1.000 & .286 & .250 & .333 \\
Under & DRC           & 30 &  7 & .378 & .233 & 1.000 & .333 & .207 & .857 \\
Under & Myanmar       & 30 &  3 & .182 & .100 & 1.000 & .182 & .100 & 1.000 \\
Under & Somalia       & 30 &  7 & .378 & .233 & 1.000 & .125 & .111 & .143 \\
Under & South Sudan   & 30 &  9 & .462 & .300 & 1.000 & .364 & .250 & .667 \\
Under & Yemen         & 30 &  8 & .421 & .267 & 1.000 & .316 & .273 & .375 \\
\cmidrule(lr){1-10}
\textit{Under (tier)} & & \textit{210} & \textit{53} & \textit{.403} & \textit{.252} & \textit{1.000} & \textit{.323} & \textit{.221} & \textit{.604} \\
\midrule
Mid & Colombia        & 30 &  6 & .000 & .000 & .000 & .000 & .000 & .000 \\
Mid & Ethiopia        & 30 & 11 & .486 & .346 & .818 & .000 & .000 & .000 \\
Mid & Haiti           & 30 & 12 & .571 & .400 & 1.000 & .571 & .400 & 1.000 \\
Mid & Lebanon         & 30 &  8 & .000 & .000 & .000 & .000 & .000 & .000 \\
Mid & Mali            & 30 &  9 & .462 & .300 & 1.000 & .462 & .300 & 1.000 \\
Mid & Mexico          & 30 &  0 & .000 & .000 & .000 & .000 & .000 & .000 \\
Mid & Nigeria         & 30 &  7 & .389 & .241 & 1.000 & .545 & .750 & .429 \\
Mid & Palestine       & 30 &  7 & .378 & .233 & 1.000 & .296 & .200 & .571 \\
Mid & Sudan           & 30 &  6 & .333 & .200 & 1.000 & .333 & .200 & 1.000 \\
Mid & Syria           & 30 &  5 & .364 & .333 & .400 & .000 & .000 & .000 \\
Mid & Ukraine         & 30 &  3 & .182 & .100 & 1.000 & .235 & .143 & .667 \\
\cmidrule(lr){1-10}
\textit{Mid (tier)}   & & \textit{330} & \textit{74} & \textit{.386} & \textit{.261} & \textit{.743} & \textit{.356} & \textit{.281} & \textit{.486} \\
\midrule
Over & Belarus         & 30 &  0 & .000 & .000 & .000 & .000 & .000 & .000 \\
Over & Iran            & 30 &  4 & .222 & .200 & .250 & .000 & .000 & .000 \\
Over & Israel          & 30 &  1 & .065 & .033 & 1.000 & .000 & .000 & .000 \\
Over & Poland          & 30 &  0 & .000 & .000 & .000 & .000 & .000 & .000 \\
\cmidrule(lr){1-10}
\textit{Over (tier)}  & & \textit{120} & \textit{5} & \textit{.100} & \textit{.057} & \textit{.400} & \textit{.000} & \textit{.000} & \textit{.000} \\
\bottomrule
\end{tabular}
\end{adjustbox}

\end{center}

\clearpage

\begin{center}
\footnotesize
\setlength{\tabcolsep}{3.5pt}
\renewcommand{\arraystretch}{1.02}

\captionof{table}{Per-country escalation forecasting results on the 660-case held-out test set: evidence-augmented condition. N\textsuperscript{+} = positive cases. Italicized rows show tier-level aggregates matching Tables~2 and~3.}
\label{tab:per_country_evidence}

\vspace{4pt}

\begin{adjustbox}{max width=\textwidth,center}
\begin{tabular}{@{}llrrrrrrrr@{}}
\toprule
& & & & \multicolumn{3}{c}{Llama-3.3-70B} & \multicolumn{3}{c}{GPT-4o} \\
\cmidrule(lr){5-7} \cmidrule(lr){8-10}
Tier & Country & N & N\textsuperscript{+} & F1 & P & R & F1 & P & R \\
\midrule
Under & Burkina Faso  & 30 & 10 & .583 & .500 & .700 & .250 & .333 & .200 \\
Under & CAR           & 30 &  9 & .571 & .500 & .667 & .133 & .167 & .111 \\
Under & DRC           & 30 &  7 & .000 & .000 & .000 & .000 & .000 & .000 \\
Under & Myanmar       & 30 &  3 & .500 & .333 & 1.000 & .750 & .600 & 1.000 \\
Under & Somalia       & 30 &  7 & .125 & .111 & .143 & .000 & .000 & .000 \\
Under & South Sudan   & 30 &  9 & .235 & .250 & .222 & .250 & .286 & .222 \\
Under & Yemen         & 30 &  8 & .222 & .200 & .250 & .000 & .000 & .000 \\
\cmidrule(lr){1-10}
\textit{Under (tier)} & & \textit{210} & \textit{53} & \textit{.342} & \textit{.300} & \textit{.396} & \textit{.167} & \textit{.186} & \textit{.151} \\
\midrule
Mid & Colombia        & 30 &  6 & .167 & .167 & .167 & .000 & .000 & .000 \\
Mid & Ethiopia        & 30 & 11 & .381 & .400 & .364 & .333 & .429 & .273 \\
Mid & Haiti           & 30 & 12 & .333 & .333 & .333 & .333 & .333 & .333 \\
Mid & Lebanon         & 30 &  8 & .125 & .125 & .125 & .125 & .125 & .125 \\
Mid & Mali            & 30 &  9 & .105 & .100 & .111 & .118 & .125 & .111 \\
Mid & Mexico          & 30 &  0 & .000 & .000 & .000 & .000 & .000 & .000 \\
Mid & Nigeria         & 30 &  7 & .316 & .250 & .429 & .133 & .125 & .143 \\
Mid & Palestine       & 30 &  7 & .118 & .100 & .143 & .143 & .143 & .143 \\
Mid & Sudan           & 30 &  6 & .267 & .222 & .333 & .235 & .182 & .333 \\
Mid & Syria           & 30 &  5 & .182 & .167 & .200 & .000 & .000 & .000 \\
Mid & Ukraine         & 30 &  3 & .000 & .000 & .000 & .000 & .000 & .000 \\
\cmidrule(lr){1-10}
\textit{Mid (tier)}   & & \textit{330} & \textit{74} & \textit{.212} & \textit{.188} & \textit{.243} & \textit{.165} & \textit{.155} & \textit{.176} \\
\midrule
Over & Belarus         & 30 &  0 & .000 & .000 & .000 & .000 & .000 & .000 \\
Over & Iran            & 30 &  4 & .250 & .167 & .500 & .375 & .250 & .750 \\
Over & Israel          & 30 &  1 & .167 & .091 & 1.000 & .000 & .000 & .000 \\
Over & Poland          & 30 &  0 & .000 & .000 & .000 & .000 & .000 & .000 \\
\cmidrule(lr){1-10}
\textit{Over (tier)}  & & \textit{120} & \textit{5} & \textit{.150} & \textit{.086} & \textit{.600} & \textit{.188} & \textit{.111} & \textit{.600} \\
\bottomrule
\end{tabular}
\end{adjustbox}

\end{center}

\end{document}